\documentclass[12pt]{article}
\usepackage{epsfig}
\usepackage[usenames]{color}
\usepackage{amsmath}
\usepackage{amsfonts}
\usepackage{amssymb}
\usepackage{mathrsfs} 
\usepackage{commath}  

\newcommand{\fsl}[1]{\ensuremath{\mathrlap{\!\not{\phantom{#1}}}#1}}
\usepackage{hyperref}

\usepackage{graphicx}
\usepackage{dcolumn}
\usepackage{bm}
\usepackage{amsmath}
\usepackage[export]{adjustbox}
\usepackage{mathtools}
\allowdisplaybreaks

\renewcommand{\theequation}{\mbox{\arabic{section}.\arabic{equation}}}
\newcounter{Roman}
\addtocounter{Roman}{1}

\newcommand{\beq}{\begin{equation}}
\newcommand{\eeq}{\end{equation}}
\newcommand{\bea}{\begin{eqnarray}}
\newcommand{\eea}{\end{eqnarray}}
\newcommand{\bml}{\begin{multline}}
\newcommand{\bal}{\begin{align}}

\newcounter{saveeqn}

\begin{document}

\begin{center}{\Large\bf  Searching for Gravity Without a Metric}\\[2cm] 
{Lukas W. Lindwasser\footnote{\sf e-mail: lukaslindwasser@physics.ucla.edu}   and 
E. T. Tomboulis\footnote{\sf e-mail: tomboulis@physics.ucla.edu}
}\\
{\em Mani L. Bhaumik Institute for Theoretical Physics\\
Department of Physics and Astronomy, UCLA, Los Angeles, 
CA 90095-1547} 
\end{center}
\vspace{1.0cm}

\begin{center}{\Large\bf Abstract}\end{center}
It has been explicitly shown how a theory with global $GL(d,\mathbb{R})$ coordinate (affine) invariance which is spontaneously broken down to its Lorentz subgroup will have as its Goldstone fields enough degrees of freedom to create a metric and a covariant derivative. Such a theory would constitute an effective theory of gravity. So far however, no explicit model has been found which exhibits this symmetry breaking pattern, mainly due to the difficulty of even writing down a $GL(d,\mathbb{R})$ invariant action in the absence of a metric. In this paper we explicitly construct an affine generalization of the Dirac action employing infinite dimensional spinorial representations of the group. This implies that it is built from an infinite number of spinor Lorentz multiplets. We introduce a systematic procedure for constructing   
$GL(d,\mathbb{R})$ and $SL(d,\mathbb{R})$ invariant interaction terms to obtain quite general interacting models. Such models have order operators whose expectation value can break affine symmetry to Poincar\'{e} symmetry. We discuss possible interactions and mechanisms for this symmetry breaking to occur, which would provide a dynamical explanation of the Lorentzian signature of spacetime.

\vfill
\pagebreak

\section{Introduction and Summary \label{sec1}}
\setcounter{equation}{0}
\setcounter{Roman}{0}

An often taken for granted observation is that physical observables in the absence of gravity are invariant under global $SO(3,1)$ spacetime coordinate transformations. Although this invariance has of course  been exhaustively confirmed experimentally, one is left wondering why flat space physics is $SO(3,1)$ invariant and not invariant under, for instance, $SO(4)$, $SO(2,2)$ or some other group of coordinate transformations. In other words, is there some 
dynamical explanation of the Lorentzian signature of spacetime?   

One possible explanation is the spontaneous symmetry breaking of a larger spacetime symmetry group.
Just as the laws governing the dynamics of a ferromagnet respect $SO(3)$ rotation invariance but at low enough temperatures the dipole moments tend to align, spontaneously breaking $SO(3)\to SO(2)$, we could imagine the laws governing the fundamental particle interactions respecting a larger coordinate symmetry group $G$ which spontaneously breaks down to $SO(3,1)$. To keep different spacetime symmetry groups, such as  $SO(3,1)$ and $SO(4)$,  on equal footing, it is necessary to assume invariance under at least global $SL(4,\mathbb{R})\subset G$ coordinate transformations, with $GL(4,\mathbb{R})$ being the largest possible group.\footnote{Note that in this discussion we restrict ourselves to global symmetry groups, and we consider physical spacetime to be four dimensional.} We will choose $GL(4,\mathbb{R})$ as our starting point and consider its spontaneous breaking to $SO(3,1)$, or more generally, any of  its subgroups.  

A theory invariant under global $GL(4,\mathbb{R})$ coordinate transformations (plus translations) lives on an affine space, with no invariant notion of distance or volume, the only invariant concept being whether two affine subspaces intersect. Recently though, it has been explicitly shown that a theory that has global $GL(4,\mathbb{R})$ (or $SL(4,\mathbb{R})$) coordinate invariance which is spontaneously broken down to $SO(3,1)$ will have as its Goldstone fields enough degrees of freedom to create a metric and a covariant derivative \cite{Tomboulis:2011qh} (for earlier work see \cite{Ogievet, Borisov, ISHAM197198}). The global $GL(4,\mathbb{R})$ symmetry is realized nonlinearly through these ingredients. Subsequently, the effective theory is found to be generally covariant, with the Einstein-Hilbert term being the low energy leading term in the effective action. This scenario therefore would not only give a satisfactory explanation for why spacetime has  $SO(3,1)$ as its symmetry group, but would  also provide a description of gravity which is inherently effective and renders the perturbative nonrenormalizability of gravity a nonissue. 

This idea is related to early work in which the photon and graviton are seen as Goldstone bosons of Lorentz symmetry breaking \cite{Bjorken} (and in more recent work \cite{LorentzBreaking}). Because the Lorentz symmetry is broken, a theory of this sort would necessarily differ from Maxwell theory and GR. In contrast, the aim here is to obtain the graviton as a Goldstone boson, while maintaining the full Lorentz group, via spontaneous breaking of affine spacetime symmetry as described above. The photon too can be realized as a Goldstone boson without sacrificing Lorentz symmetry via continuous one-form global symmetry breaking \cite{higherform}.

The result obtained in \cite{Tomboulis:2011qh} assumes only the symmetry breaking pattern, and nothing else about the details of the underlying theory. While powerful in its generality, this result is currently lacking in the existence of any concrete examples that exhibit this symmetry breaking pattern. The aim of this paper is to begin the search for a concrete example.

If one starts looking for global $GL(4,\mathbb{R})$ invariant actions, they will quickly conclude that any such action built out of fields that are finite representations of $GL(4,\mathbb{R})$ will have trivial dynamics. For instance, a field $\Phi(x)$ which transforms like $\Phi(x)\to\sqrt{|\det M^{-1}|}\Phi(M^{-1}x)$ under the affine coordinate change $x\to Mx$ has the invariant action $\int d^4x\frac{1}{2}\Phi^2(x)$. This is an auxiliary field whose correlation function $\langle\Phi(x)\Phi(y)\rangle=\delta^4(x-y)$ is trivial (notice that the delta function appropriately measures whether the two points $x$ and $y$ intersect). Similarly, one can only construct topological actions for vector field representations. Tensor representations will have similar limitations. This leaves infinite dimensional representations of $GL(4,\mathbb{R})$ as our remaining option. It is possible that something nontrivial emerges after incorporating an infinite number of fields. Because the possibility of a dynamically generated metric is so interesting, we will continue searching.

A decent body of literature \cite{etde_5883936, doi:10.1063/1.526758, Sijacki:1986af, doi:10.1063/1.526646, Kirsch_2002, ija_ki_2004} exists on infinite dimensional representations of the affine group, mainly in an attempt to define spinors on a curved manifold. A theorem which goes back to Cartan \cite{cartan1981theory} states that there exists no finite dimensional spinor representations of $GL(d,\mathbb{R})$. Because of this, when spinors are lifted onto a curved background, they are usually defined to only transform with respect to local Lorentz transformations. An affine spinor (affinor?), which is necessarily infinite dimensional, enjoys transforming under the full diffeomorphism group on a curved manifold. Affine spinors can be decomposed in terms of their Lorentz (or Euclidean) spin content, which will generally be an infinite tower of all possible half-integer spins. Because half-integer spin representations can be used to construct integer spin representations, we view the affine spinor representations as in some sense more fundamental than their integer spin cousins. Consequently, we will focus on constructing a nontrivial affine theory which spontaneously breaks down to its Lorentz subgroup out of these. Infinite dimensional tensorial affine theories may be constructed from the affine spinor if one wishes.

This paper is organized as follows. In section \ref{sec2} we will give a brief review of the result detailed in \cite{Tomboulis:2011qh}. In section \ref{sec3}, we will use an infinite dimensional spinor representation of $GL(d,\mathbb{R})$ to construct an affine invariant Dirac action in any dimension $d$. This spinor representation affords an order operator which if nonzero can spell the breakdown of affine invariance down to its Lorentz subgroup. In this construction, we will find it useful to embed our $GL(d,\mathbb{R})$ representation within $SL(d+1,\mathbb{R})$, so that an affine generalization of the gamma matrix $\Gamma^{\mu}$ exists. In section \ref{sec4}, we restrict ourselves to $d=3$ essentially for technical  reasons explained in the text, i.e., spinor representations of $SL(n,\mathbb{R})$ have only been written down in exact detail for $n\leq 4$ \cite{Sijacki:1986af}, and so we give an explicit construction of the action in this case. In section \ref{sec5}, we will show how to introduce interactions, which is of course necessary for the goal of studying spontaneous symmetry breaking. Section \ref{sec6} contains our conclusions. 

\section{\texorpdfstring{Gravity as the Effective Theory of $GL(d,\mathbb{R})\to SO(d-1,1)$ Symmetry Breaking}{Gravity as the Effective Theory of GL(d,R) to SO(d-1,1) Symmetry Breaking} \label{sec2}} 
\setcounter{equation}{0}
\setcounter{Roman}{0}  
The $d$ dimensional Affine algebra consists of generators $Q^{\alpha}_{\;\;\beta}$ and $P_{\gamma}$, which generate rotations, boosts, dilatations, and translations in all coordinate directions, respectively. Here $\alpha,\beta,\gamma=1,\cdots, d$. The algebra reads
\begin{align} \label{eq1:1}
&[Q^{\alpha}_{\;\;\beta},Q^{\gamma}_{\;\;\delta}]=i\delta^{\alpha}_{\;\;\delta}Q^{\gamma}_{\;\;\beta}-i\delta^{\gamma}_{\;\;\beta}Q^{\alpha}_{\;\;\delta},\\ \label{eq1:2}
&[Q^{\alpha}_{\;\;\beta},P_{\gamma}]=i\delta^{\alpha}_{\;\;\gamma}P_{\beta},\\ \label{eq1:3}
&[P_{\alpha},P_{\beta}]=0.
\end{align}

Here, indices that are raised are contravariant, transforming like a coordinate $x^{\alpha}$, and lowered indices are covariant, transforming like a derivative $\partial_{\alpha}$. 

Frequently, it is useful to parametrize these generators with respect to some subgroup which leaves some constant metric $g_{\alpha\beta}$ invariant (here $g_{\alpha\beta}$ could be the Euclidean metric $\delta_{\alpha\beta}$ or the Minkowski metric $\eta_{\alpha\beta}$ etc.). To do this, we express $Q^{\alpha}_{\;\;\beta}=\frac{1}{2}(J^{\alpha}_{\;\;\beta}+T^{\alpha}_{\;\;\beta})$, where $J_{\alpha\beta}\equiv g_{\alpha\gamma}J^{\gamma}_{\;\;\beta}$ is antisymmetric and $T_{\alpha\beta}\equiv g_{\alpha\gamma}T^{\gamma}_{\;\;\beta}$ is symmetric (and traceless when restricting to the special linear group). Here $J_{\alpha\beta}$ generates the subgroup that leaves the metric $g$ invariant, and $T_{\alpha\beta}$ generates the ``rest" of the coordinate transformations. In terms of these generators, the algebra reads
\begin{align} \label{eq2:1}
&[J_{\alpha\beta},J_{\gamma\delta}]=i(g_{\alpha\gamma}J_{\beta\delta}-g_{\alpha\delta}J_{\beta\gamma}-g_{\beta\gamma}J_{\alpha\delta}+g_{\beta\delta}J_{\alpha\gamma}),\\ \label{eq2:2}
&[J_{\alpha\beta},T_{\gamma\delta}]=i(g_{\alpha\gamma}T_{\beta\delta}+g_{\alpha\delta}T_{\beta\gamma}-g_{\beta\gamma}T_{\alpha\delta}-g_{\beta\delta}T_{\alpha\gamma}),\\ \label{eq2:3}
&[T_{\alpha\beta},T_{\gamma\delta}]=-i(g_{\alpha\gamma}J_{\beta\delta}+g_{\alpha\delta}J_{\beta\gamma}+g_{\beta\gamma}J_{\alpha\delta}+g_{\beta\delta}J_{\alpha\gamma}),\\ \label{eq2:4}
&[J_{\alpha\beta},P_{\gamma}]=i(g_{\alpha\gamma}P_{\beta}-g_{\beta\gamma}P_{\alpha}),\\ \label{eq2:5}
&[T_{\alpha\beta},P_{\gamma}]=i(g_{\alpha\gamma}P_{\beta}+g_{\beta\gamma}P_{\alpha}),\\ \label{eq2:6}
&[P_{\alpha},P_{\beta}]=0.
\end{align}
For our purposes, we will take $g_{\alpha\beta}=\eta_{\alpha\beta}=\text{diag}(-1,+1,\cdots,+1)$. Note now in this description, lowered indices transform covariantly only under the Lorentz group, and indices raised by $\eta^{\alpha\beta}$ contravariantly only under the Lorentz group. With this parametrization, we choose the indices to run from $0$ to $d-1$.
\subsection{\texorpdfstring{Coset Construction of $GL(d,\mathbb{R})\to SO(d-1,1)$}{Coset Construction of GL(d,R) to SO(d-1,1)}} \label{sec2:1}

We now assume spontaneous breaking of the global $GL(d,\mathbb{R})=G$ to its $SO(d-1,1)=H$ subgroup. Such a theory is best described by the now standard coset construction \cite{PhysRev.177.2239} and subsequent modifications for spacetime symmetries \cite{Nonlinear} which identifies field content that transforms linearly under the unbroken group $H$, but nonlinearly under the full group $G$. This construction allows for any $H$ invariant action built out of these fields to be invariant under $G$. For internal symmetries, one may start by noting that any field $\psi$ which transforms linearly under $G$ may be split into two factors
\begin{equation} \label{eq3}
\psi(x)=\gamma(\xi)\tilde{\psi}(x), \quad \gamma(\xi)=\exp\big(\frac{i}{2}\xi(x)\cdot T\big)\in G/H.
\end{equation}
Where the notation $\xi\cdot T \equiv \xi_{\alpha\beta}T^{\alpha\beta}$ was used, and $T^{\alpha\beta}$ are the broken generators. To accommodate for spacetime symmetries it is useful to further parametrize
\begin{equation} \label{eq4}
\psi(x)=\Gamma(\xi)\tilde{\psi}=\exp\big(-ix^{\mu}P_{\mu}\big)\exp\big(\frac{i}{2}\xi(x)\cdot T\big)\tilde{\psi}(x).
\end{equation}
Under the group action $\psi(x)\to\psi'(x')=g\psi(x')$ with $g\in G$ and $x'^{\mu}=M^{\mu}_{\;\;\nu}(g)x^{\nu}$, we see that 
\begin{equation} \label{eq5}
g\exp\big(-ix'^{\mu}P_{\mu}\big)\exp\big(\frac{i}{2}\xi(x')\cdot T\big)=\exp\big(-i(M^{-1}x')^{\mu}P_{\mu}\big)\exp\big(\frac{i}{2}\xi'(x')\cdot T\big)\exp\big(\frac{i}{2}u(\xi,g)\cdot J\big)
\end{equation}
defining a nonlinear realization of $GL(d,\mathbb{R})$ coordinate transformations on the fields $\xi(x)\to\xi'(x')=\xi'(x)$ and $\tilde{\psi}(x)\to\tilde{\psi}'(x')=\exp\big(\frac{i}{2}u(\xi,g)\cdot J\big)\tilde{\psi}(x)$. When $g\in H$, these transformations become linear. 

To construct an effective Lagrangian out of these one also needs an appropriate notion of derivation on these fields. This can be obtained by considering the Maurer-Cartan form $\Gamma^{-1}d\Gamma$
\begin{equation} \label{eq6}
\Gamma^{-1}d\Gamma=i\hat{\omega}^{\alpha}P_{\alpha}+\frac{i}{2}D_{\alpha\beta}T^{\alpha\beta}+\frac{i}{2}\omega_{\alpha\beta}J^{\alpha\beta}.
\end{equation}

A calculation \cite{Tomboulis:2011qh} shows
\begin{equation} \label{eq7}
\hat{\omega}^{\alpha}=dx^{\mu}e_{\mu}^{\;\;\alpha},\quad D_{\alpha\beta}=\frac{1}{2}\{e^{-1},de\}_{\alpha\beta},\quad \omega_{\alpha\beta}=\frac{1}{2}[e^{-1},de]_{\alpha\beta}
\end{equation}
where $e$ is a symmetric invertible matrix
\begin{equation} \label{eq8}
e_{\alpha\beta}\equiv(\exp(\xi))_{\alpha\beta}.
\end{equation}

$\hat{\omega}^{\alpha}$ and $D_{\alpha\beta}$ transform homogeneously while $\omega_{\alpha\beta}$ transforms inhomogeneously (like a gauge field). $\hat{\omega}^{\alpha}$ provides a basis of 1-forms which we will use to define the covariant derivative:
\begin{equation} \label{eq9}
D_{\mu}\tilde{\psi}=e^{-1\alpha}_{\mu}\partial_{\alpha}\tilde{\psi}+\frac{i}{2}\omega_{\mu\alpha\beta}J^{\alpha\beta}\tilde{\psi}.
\end{equation}
Where $\omega_{\mu\alpha\beta}$ is a particular combination of Goldstone `gauge field' $\omega_{\alpha\beta}=\hat{\omega}^{\mu}\tilde{\omega}_{\mu\alpha\beta}$ and covariant derivative $D_{\alpha\beta}=\hat{\omega}^{\mu}\mathcal{D}_{\mu\alpha\beta}$ as written in the $\hat{\omega}$ basis. Note that $\hat{\omega}$ is dual to the vector basis $\hat{e}_{\alpha}=e_{\alpha}^{-1\mu}\partial_{\mu}$ so that $\langle\hat{\omega}^{\alpha},\hat{e}_{\beta}\rangle=\delta^{\alpha}_{\;\;\beta}$.

Any Lagrangian which is $H$ (Poincar\'{e}) invariant and written in terms of the building blocks $\mathcal{D}$, $\tilde{\psi}$, $D\tilde{\psi}$ etc. is automatically $G$ (Affine) invariant. This provides a general way to describe an effective theory of $GL(d,\mathbb{R})\to SO(d-1,1)$ breaking. A yet more powerful construction of the effective theory, which better describes the physical field content is to use the ingredients we have so far developed to pass over to fields which transform linearly with the full group $G$, rather than only with $H$. For instance, a covariant vector under $H$ with components $v_{\alpha}$ may be converted to a covariant $G$ vector $V_{\mu}$ via
\begin{equation} \label{eq10}
V_{\mu}=e_{\mu}^{\;\;\alpha}v_{\alpha}.
\end{equation}
Indeed, one can readily check that $V_{\mu}$ transforms linearly under $G$:
\begin{equation} \label{eq11}
V'(x')=(\exp(\xi'(x')))v'(x')=(\exp(\xi'(x')))\Lambda(u(\xi,g))v(x)=M^{-1}(g)(\exp(\xi(x)))v(x).
\end{equation}
In fact, the same passage can be made for any tensor representation of $H$
\begin{equation} \label{eq12}
\tilde{\Psi}^{\mu_1\cdots\mu_l}_{\nu_1\cdots\nu_k}=e_{\nu_1}^{\;\;\alpha_1}\cdots e_{\nu_k}^{\;\;\alpha_k}e_{\beta_1}^{-1\mu_1}\cdots e_{\beta_l}^{-1\mu_l}\tilde{\psi}^{\beta_1\cdots\beta_l}_{\alpha_1\cdots\alpha_k}.
\end{equation}
This is equivalent to the basis change $\{\hat{\omega},\hat{e}\}\to\{dx,\partial\}$. Notice that this cannot be done for spinor representations of $H$, and they will have to remain in the $\{\hat{\omega},\hat{e}\}$ basis (as will become clear this is in fact natural). Of particular importance for this discussion is when this passage is applied to the $H$ invariant tensor $\eta_{\alpha\beta}$:
\begin{equation} \label{eq13}
g_{\mu\nu}=e_{\mu}^{\;\;\alpha}e_{\nu}^{\;\;\beta}\eta_{\alpha\beta}.
\end{equation}
This symmetric rank two tensor has an inverse $g^{\mu\nu}$ such that $g^{\mu\lambda}g_{\lambda\nu}=\delta^{\mu}_{\;\;\nu}$, defined similarly to be $g^{\mu\nu}=e_{\alpha}^{-1\mu}e_{\beta}^{-1\nu}\eta^{\alpha\beta}$. This clearly has the form of a metric, built solely out of the Goldstone fields $\xi$, with vielbein $e_{\mu}^{\;\;\alpha}$.

One may further show \cite{Tomboulis:2011qh} that the covariant derivative appropriately defines a metric compatible connection $\Gamma^{\mu}_{\;\;\nu\lambda}$ under this basis change, such that 
\begin{equation} \label{eq14}
\Gamma^{\mu}_{\;\;\nu\lambda}=\frac{1}{2}g^{\mu\kappa}(g_{\kappa\nu ,\lambda}+g_{\kappa\lambda , \nu}-g_{\nu\lambda , \kappa})
\end{equation}
and through this may further define a Riemann curvature tensor $R^{\mu}_{\;\;\nu\rho\sigma}$. With these new ingredients it is clear that any effective Lagrangian will be constructed via the basic framework of General Relativity, the Einstein-Hilbert action plus a cosmological term being the unique lowest energy term to add for the Goldstone fields.

One may ask where the general coordinate invariance arises out of a theory which only assumed global $GL(d,\mathbb{R})$ coordinate invariance. General coordinate invariance appears from the fact that any $G$ invariant expression built out of the ingredients developed remains invariant under transformations $M^{\mu}_{\;\;\nu}(g)\in G$ which have been made spacetime dependent, i.e. a transformation $\partial x'^{\mu}/\partial x^{\nu}$ for any differentiable $x'(x)$. This general coordinate invariance ultimately reduces the Goldstone field degrees of freedom to that of a graviton. In this sense, the graviton is quite literally the Goldstone particle of global $GL(d,\mathbb{R})\to SO(d-1,1)$ symmetry breaking.

\section{Spinor Representations of $GL(d,\mathbb{R})$ and the Affine Dirac Action \label{sec3}}
\setcounter{equation}{0}
\setcounter{Roman}{0}

The result of the previous section is interesting in that it does not assume any details of the underlying theory other than the symmetry breaking pattern. This begs the question of whether this scenario is realizable. To date, there are no explicit examples of this symmetry breaking. The obstacles ahead for constructing such a theory are obvious. Any $GL(d,\mathbb{R})$ action you might wish to construct out of finite dimensional representations without the use of a metric (we want the graviton to be produced dynamically through this symmetry breaking process) will be either topological or auxiliary as we saw in section \ref{sec1}. We must therefore resort to looking at infinite dimensional representations of $GL(d,\mathbb{R})$ to construct such a theory.

A decent body of literature \cite{etde_5883936, doi:10.1063/1.526758, Sijacki:1986af, doi:10.1063/1.526646, Kirsch_2002, ija_ki_2004} exists on infinite dimensional representations of the affine group, mainly in an attempt to define spinors on a curved manifold. A theorem which goes back to Cartan \cite{cartan1981theory} states that there exists no finite dimensional spinor representations of $GL(d,\mathbb{R})$. Because of this, when spinors are lifted onto a curved background, they are usually defined to only transform with respect to local Lorentz transformations. An affine spinor, which is necessarily infinite dimensional, enjoys transforming under the full diffeomorphism group on a curved manifold.

Spinor representations of $GL(d,\mathbb{R})$ are obtained through its double cover $\overline{GL(d,\mathbb{R})}$. Under the Iwasawa decomposition $GL(d,\mathbb{R})$ may be split into three factors $KAN$, where $A$ is the maximal abelian subgroup of positive diagonal matrices, $N$ is the nilpotent subgroup of upper triangular matrices with 1's along the diagonal, and $K$ is the maximally compact subgroup $O(d)$ (this is essentially the $QR$ decomposition of square matrices). The factors $A$, $N$ are connected, so the double cover $\overline{GL(d,\mathbb{R})}$ is obtained by replacing $K$ with its double cover $\overline{K}$, i.e. $O(d)\to \text{Pin}(d)$.\footnote{This double covering is not unique, as there are two non-isomorphic pin groups $\text{Pin}_{\pm}$.} $GL(d,\mathbb{R})$ spinors may therefore be built out of the standard $O(d)$ spinors. The double cover may similarly be defined as replacing $O(d-1,1)$ with its double cover, which is what we do in what follows. When the context is clear, we will refer to the double cover of $GL(d,\mathbb{R})$ as $GL(d,\mathbb{R})$ itself.

Assuming we have found such a spinor representation of $GL(d,\mathbb{R})$, it is possible to construct an affine invariant action out of a field $\Psi(x)$ in complete analogy to the Dirac action. In particular, if we assume that under the coordinate transformation $x'^{\mu}=M^{\mu}_{\;\;\nu}x^{\nu}$,
\begin{equation} \label{eq15}
\Psi'(x')=\sqrt{|\det{M^{-1}}|}U(M)\Psi(M^{-1}x')
\end{equation}
with $U(M)$ some representation, and further postulate the existence of a constant affine vector operator $\Gamma^{\mu}$ with the transformation property
\begin{equation} \label{eq16}
U^{-1}(M)\Gamma^{\mu}U(M)=M^{\mu}_{\;\;\nu}\Gamma^{\nu},
\end{equation}
and the existence of an affine invariant inner product\footnote{This will be positive definite in the case of a unitary representation, but this is not required.} $(\cdot,\cdot)$ such that $(Uf,Ug)=(f,g)$, then the following action is affine invariant
\begin{equation} \label{eq17}
S=-\int d^dx(\Psi,\Gamma^{\mu}\partial_{\mu}\Psi).
\end{equation}

It remains to find a representation which admits the vector $\Gamma^{\mu}$ and inner product $(\cdot,\cdot)$ (see subsection \ref{sec3:1} and section \ref{sec4} below for the explicit details). Before we do this, it is important to first make sure that a theory with these objects, after incorporating interactions (see section \ref{sec5}), has the ability to break the symmetry. By this we mean there must be some order operator whose expectation value remains invariant under some subgroup $H$ but not invariant under $GL(d,\mathbb{R})/H$ transformations. The theory so far described has many such order operators, the simplest one being
\begin{equation} \label{eq18}
(\Psi,\Gamma^{\mu}\Gamma^{\nu}\Psi).
\end{equation}
This will, for the representations we consider in subsection \ref{sec3:1} and section \ref{sec4}, always be a real symmetric rank two tensor, whose nonzero expectation value $\langle(\Psi,\Gamma^{\mu}\Gamma^{\nu}\Psi)\rangle=\bar{g}^{\mu\nu}$ will break $GL(d,\mathbb{R})$ to some subgroup $H$ that leaves this constant tensor $\bar{g}^{\mu\nu}$ invariant. For instance if $\bar{g}^{\mu\nu}=\eta^{\mu\nu}$, the symmetry is broken to $SO(d-1,1)$. If instead $\bar{g}^{\mu\nu}=\delta^{\mu\nu}$, the symmetry would break to $SO(d)$. One can always find a $GL(d,\mathbb{R})$ transformation which makes a real 
$\bar{g}^{\mu\nu}$ a diagonal matrix of $1$'s, $-1$'s and $0$'s. Excluding the possibility of 0's, which would make $\bar{g}^{\mu\nu}$ non-invertible, the number of possible broken phases endowed with a metric that this order operator can distinguish is $\lfloor\frac{d}{2}+1\rfloor$,\footnote{Here $\lfloor x\rfloor$ is the floor function, which rounds the real valued $x$ down to the greatest integer $n$ such that $n\leq x$. Similarly, the ceiling function $\lceil x\rceil$ rounds $x$ up to the smallest integer $n$ such that $n\geq x$.} corresponding to the unbroken subgroups $SO(p,q)$ such that $p\geq q \geq 0$ and $p+q=d$. Which subgroup a theory breaks down to will depend on the interactions, range of couplings, and choice of representation.

\subsection{Embedding of $GL(d,\mathbb{R})$ in $SL(d+1,\mathbb{R})$ and the Deunitarizing Automorphism} \label{sec3:1}
\setcounter{Roman}{0}

In contrast to spinor representations of the Lorentz group, the existence of a vector operator and inner product in a spinor representation of $GL(d,\mathbb{R})$ are far from guaranteed. In fact, it has been shown \cite{doi:10.1063/1.526646} that for $d=4$ the simplest, multiplicity-free representations (representations which consist of at most one of each irreducible $SO(4)$ $(j_1,j_2)$ multiplet) you might think to use \textit{do not} admit vector operators.

If one embeds $GL(d,\mathbb{R})$ within $SL(d+1,\mathbb{R})$ and uses representations of the larger group, these \textit{will} admit vector operators. To see this, note that if $Q^{\alpha}_{\;\;\beta}$ with $\alpha,\beta=1,\dots,d+1$ are the generators of $SL(d+1,\mathbb{R})$, and $Q^{\mu}_{\;\;\nu}$ with $\mu,\nu=1,\dots,d$ are the generators of the $GL(d,\mathbb{R})$ subgroup, equations (\ref{eq1:1}) show that $\Gamma^{\mu}\equiv Q^{\mu}_{\;\;d+1}$ and $\Omega_{\mu}\equiv Q^{d+1}_{\;\;\;\;\;\;\mu}$ are contravariant and covariant vectors, respectively, with respect to $GL(d,\mathbb{R})$. Note that $[\Gamma^{\mu},\Gamma^{\nu}]=[\Omega_{\mu},\Omega_{\nu}]=0$, and so $\Gamma^{\mu}$ do not satisfy any Clifford algebra. So although $\Gamma^{\mu}$ will play the role of $\gamma^{\mu}$ in the Lorentz Dirac action, it is strikingly different in nature. Fortunately, $[\Gamma^{\mu},\Gamma^{\nu}]=0$ ensures that the order operator (\ref{eq18}) is symmetric.

What is left is to find a spinor representation of $SL(d+1,\mathbb{R})$ which admits a $GL(d,\mathbb{R})$ invariant inner product.  To ensure the existence of an inner product, it is enough to use a unitary representation of $SL(d+1,\mathbb{R})$. Such a unitary representation acts on a vector space consisting of finite dimensional $SO(d+1)$ multiplets, and subsequently a vector $\Psi$ in this representation will be built out of an infinite number of irreducible $SO(d)$ 
multiplets.  Unitary spinor representations of $SL(n,\mathbb{R})$ have been explicitly cataloged for $n\leq 4$ \cite{doi:10.1063/1.526758, Sijacki:1986af} (cf.  \cite{doi:10.1142/S0219887811005208} for recent work on extending this to arbitrary $n$). So in the following section \ref{sec4} we will restrict ourselves to constructing affine invariant theories in dimensions $d\leq 3$. 

Let us suppose now that we have found a theory with a field $\Psi(x)$ transforming with respect to the above unitary representation, whose affine symmetry is spontaneously broken down to its Lorentz subgroup $SO(d-1,1)$. This unbroken Lorentz subgroup would then be linearly realized in the effective theory on infinite dimensional unitary Lorentz field representations (and not on an infinite number of non-unitary finite dimensional Lorentz field multiplets).

To better reflect the field content observed in nature then, one should, instead, 
find a $GL(d,\mathbb{R})$ representation which acts on a vector space consisting of finite dimensional Lorentz multiplets, so that the $GL(d,\mathbb{R})\to SO(d-1,1)$ broken effective theory is expressed in terms of an infinite number of finite dimensional non-unitary Lorentz multiplets. 
Luckily, one may still use the unitary representations above by applying a deunitarizing automorphism on the generators \cite{doi:10.1063/1.526758}:
\begin{equation} \label{eq19}
J'_{ij}=J_{ij},\quad T'_{ij}=T_{ij},\quad J'_{0k}=iT_{0k},\quad T'_{0k}=iJ_{0k},\quad T'_{00}=T_{00} \;  ,
\end{equation}
where $i,j,k$ run through spatial indices. After making this transformation the new representation will act on a vector space consisting of finite dimensional non-unitary$SO(d,1)$ (and hence $SO(d-1,1)$) multiplets. Although this representation is no longer unitary, it still has a $GL(d,\mathbb{R})$ invariant inner product. Note that the vector operators $\Gamma^{\mu}$ and $\Omega_{\mu}$ transform under this automorphism as well: $(\Gamma^0,\Gamma^{1},\cdots,\Gamma^{d-1})\to(i\Gamma^0,\Gamma^{1},\cdots,\Gamma^{d-1})$ and $(\Omega_0,\Omega_{1},\cdots,\Omega_{d-1})\to(-i\Omega_0,\Omega_{1},\cdots,\Omega_{d-1})$.

Alternatively, the affine symmetry may break to one of the other subgroups $SO(p,q)$ with $p\geq q > 1$, $p+q=d$. 
In this case, starting with a unitary representation of $SL(d+1, \mathbb{R})$ and proceeding as above, in the $GL(d,\mathbb{R})\to SO(p,q)$ effective theory the unbroken group
will be linearly realized on infinite dimensional unitary $SO(p,q)$ representations. 
It is straightforward, however, to write down the extension of the automorphism (\ref{eq19}), which, applied to the unitary representation, yields a non-unitary representation consisting of an infinite number of finite dimensional non-unitary $SO(p,q)$ representations. Which of  these symmetry breaking scenarios, i.e., which of the subgroups $SO(p,q)$, is actually realized in any particular model is of course dependent on the choice of interactions and the range of coupling strengths. Note that models with phase diagrams with more than two phases can,  in principle, occur for $d \geq 4$. 

To summarize, we have identified a general way of constructing a $GL(d,\mathbb{R})$ invariant action which has an order operator that, after incorporating interactions, can break its affine symmetry down to some subgroup $H$, of which $H=SO(d-1,1)$ is but one possibility, as long as one has on hand a unitary spinor representation of $SL(d+1,\mathbb{R})$. The current literature has only described such representations in explicit detail for $d\leq 3$. Thus, in the following section, we restrict ourselves to $d=3$.  
The important case $d=4$ is left for future work. 

\section{Explicit Construction of a $3$-dimensional Affine Dirac Action} \label{sec4}
\setcounter{equation}{0}
\setcounter{Roman}{0}
In this section, we will demonstrate explicitly the construction detailed above in the case of $d=3$. We will not go into detail here of how to construct the representation itself, but mention here that we are using the unitary irreducible $e=\frac{1}{2}$, $(0,\frac{1}{2})$ lattice representation of $\overline{SL(4,\mathbb{R})}$ (in the notation of \cite{doi:10.1063/1.526758}). Unitary representations of $\overline{SL(4,\mathbb{R})}$ are described in appendix \ref{appA}.

This unitary representation acts on the vector space built from finite dimensional multiplets of $\text{Spin}(4)$ denoted $(j_1,j_2)$, with $j_1, j_2$ standard $SU(2)$ spin labels. Specifically, it acts on an invariant lattice of points $(j_1,j_2)=(2n,2m+\frac{1}{2}),(2n+1,2m+1+\frac{1}{2})$ with $m,n$ non-negative integers and $m\geq n$. A vector $\Psi$ living on this vector space will have components $\psi^{j_1j_2}_{m_1m_2}$, and the inner product between two vectors $\Phi$ and $\Psi$ which makes this representation unitary is
\begin{equation} \label{eq20}
(\Phi,\Psi)=\sum_{j_1m_1j_2m_2}\phi^{*j_1j_2}_{m_1m_2}\kappa(j_1,j_2)\psi^{j_1j_2}_{m_1m_2},\quad \kappa(j_1,j_2)=\frac{(j_1+j_2+\frac{1}{2})!!(j_2-j_1-\frac{1}{2})!!}{(j_1+j_2-\frac{1}{2})!!(j_2-j_1-\frac{3}{2})!!}.
\end{equation}

We now move over to the deunitarized representation which is built out of finite dimensional $(j_1,j_2)$ $\text{Spin}(3,1)$ representations. Everything is essentially the same as above except a small modification of the kernel $\kappa\to -ie^{i\pi(m_1+m_2)}\kappa$ (the added factor is a $3$-dimensional parity transformation). 

Knowing the inner product and the generators which create the vector operator $\Gamma^{\mu}$ is enough to explicitly write down the action in this basis. What one should of course do in the present context is write the field in terms of its finite dimensional irreducible $d=3$ Lorentz components, since this reducible $\overline{GL(3,\mathbb{R})}$ representation is defined by how it acts on these components.

To this end, we will represent the components of $\Psi(x)$ in terms of $3$-dimensional spinor tensor fields $\psi^{n,h}_{\mu_1\cdots\mu_s}(x)$ with $s$ Lorentz tensor indices and one omitted Dirac spinor index. The labels $n,h$ indicate which $\text{Spin}(3,1)$ multiplet the component came from: $(n-h,n+h+\frac{1}{2})$, with $n=\lceil \frac{s}{2}\rceil, \lceil \frac{s}{2}\rceil + 1,\dots$ and $h=0,\dots,\lfloor\frac{s}{2}\rfloor$. The fields $\psi^{n,h}_{\mu_1\cdots\mu_s}$ are totally symmetric in their tensor indices, as well as gamma traceless $\gamma^{\mu_1}\psi^{n,h}_{\mu_1\cdots\mu_s}=0$. These two algebraic constraints on the fields are necessary to treat them as components of $\Psi$. Notice that now that we are at the level of the individual Lorentz components of the $\overline{GL(3,\mathbb{R})}$ covariant field $\Psi$, any statement made will only be manifestly covariant under the Lorentz subgroup (the gamma traceless condition is defined in terms of the standard Lorentz gamma $\gamma^{\mu}$).
\pagebreak
In terms of these ingredients, the affine invariant action is
\begin{alignat}{2} \label{eq21}
S&=-\int d^3x(\Psi,\Gamma^{\mu}\partial_{\mu}\Psi)\nonumber
\\&=- {\textstyle\sum_{s=0}^{\infty}\sum_{n=\lceil\frac{s}{2}\rceil}^{\infty}\sum_{h=0}^{\lfloor\frac{s}{2}\rfloor}}&\int d^3x\;\Bigg(\overline{\psi^{n,h}}_{\mu_1\cdots\mu_s}\Pi^{\mu_1\cdots\mu_s\lambda\nu_1\cdots\nu_s}\partial_{\lambda}\Big(\Gamma^{s,s}\psi_{\nu_1\cdots\nu_s}\Big)^{n,h}\\&&-\;\overline{\psi^{n,h}}_{\mu_1\cdots\mu_s}\Pi^{\mu_1\cdots\mu_s\lambda\nu_1\cdots\nu_{s-1}}\partial_{\lambda}\Big(\Gamma^{s,s-1}\psi_{\nu_1\cdots\nu_{s-1}}\Big)^{n,h}\nonumber\\&&-\;\overline{\psi^{n,h}}_{\mu_1\cdots\mu_s}\Pi^{\mu_1\cdots\mu_s\lambda\nu_1\cdots\nu_{s+1}}\partial_{\lambda}\Big(\Gamma^{s,s+1}\psi_{\nu_1\cdots\nu_{s+1}}\Big)^{n,h}\Bigg),\nonumber
\end{alignat}
\begin{equation}
\hspace{-6cm}\overline{\psi^{n,h}}_{\mu_1\cdots\mu_s}\equiv(-1)^s(\psi^{n,h}_{\mu_1\cdots\mu_s})^{\dagger}i\gamma^0\frac{(2n+1)!!}{(2n)!!}\frac{(2h)!!}{(2h-1)!!}
\end{equation}

Here we have introduced several new objects. First, $\overline{\psi^{n,h}}_{\mu_1\cdots\mu_s}$ is defined to incorporate the affine invariant inner product, in much the same way the `bar' notation is used in the Lorentz case. Then, we need several spinor matrix even and odd rank tensor structures $\Pi^{\mu_1\cdots\mu_s\nu_1\cdots\nu_s}$, $\Pi^{\mu_1\cdots\mu_s\lambda\nu_1\cdots\nu_s}$ respectively. These are all totally symmetric in their $\mu$ and $\nu$ indices separately, as well as gamma traceless $\gamma_{\mu_1}\Pi^{\mu_1\cdots\mu_s\nu_1\cdots\nu_s}=0$, $\Pi^{\mu_1\cdots\mu_s\nu_1\cdots\nu_s}\gamma_{\nu_1}=0$ etc. Their adjoints are $\big(\Pi^{\lambda}_s\big)^{\dagger}=\gamma_0\Pi^{\lambda}_s\gamma_0$, $\big(\Pi_s\big)^{\dagger}=-\gamma_0\Pi_s\gamma_0$, and they have the normalizations $\Pi_s\Pi_s=\Pi_s$, $\Pi_{s\lambda}\Pi^{\lambda}_s=\frac{2s+3}{2s+1}\Pi_s$. For example, $\Pi^{\mu\nu}=\frac{1}{3}(2\eta^{\mu\nu}-\gamma^{\mu\nu})$ and $\Pi^{\mu\lambda\nu}=\frac{1}{9}(4\eta^{\mu\nu}\gamma^{\lambda}-\eta^{\lambda\mu}\gamma^{\nu}-\eta^{\lambda\nu}\gamma^{\mu}+5\gamma^{\mu\lambda\nu})$. These tensor structures quickly become too long to write down, but there is an algorithm to generate all of them, explained in appendix \ref{appB}.

Finally, the objects $\Gamma^{s,s}$, $\Gamma^{s,s-1}$ and $\Gamma^{s,s+1}$ are infinite dimensional matrices acting on the $n,h$ labels. $\Gamma^{s,s}$ is Hermitian with respect to the affine invariant inner product, and $(\Gamma^{s,s+1})^{\dagger}=\Gamma^{s+1,s}$. Their explicit forms are given in appendix \ref{appB}.

The $\overline{\psi^{n,h}}_{\mu_1\cdots\mu_s}$ equations of motion are
\begin{align} \label{eq21.5}
\Gamma^{s,s}\Pi^{\mu_1\cdots\mu_s\lambda\nu_1\cdots\nu_s}\partial_{\lambda}\psi_{\nu_1\cdots\nu_s}&-\Gamma^{s,s-1}\Pi^{\mu_1\cdots\mu_s\lambda\nu_1\cdots\nu_{s-1}}\partial_{\lambda}\psi_{\nu_1\cdots\nu_{s-1}}\nonumber\\&-\Gamma^{s,s+1}\Pi^{\mu_1\cdots\mu_s\lambda\nu_1\cdots\nu_{s+1}}\partial_{\lambda}\psi_{\nu_1\cdots\nu_{s+1}}=0.
\end{align}
These equations are consistent in the sense that they maintain the gamma traceless and symmetric constraints on the fields $\gamma^{\mu_1}\psi^{n,h}_{\mu_1\cdots\mu_s}=0$.

\section{Incorporating Interactions \label{sec5}}
\setcounter{equation}{0}
\setcounter{Roman}{0}

So far we have constructed a free affine invariant theory out of an infinite dimensional representation. It is worth discussing what using an infinite dimensional representation has bought us in terms of finding a theory whose affine symmetry spontaneously breaks. As previously discussed, any finite dimensional representation of $GL(d,\mathbb{R})$ will give trivial correlation functions. Indeed, any propagator $G(x)=\langle\Phi(x)\Phi(0)\rangle$ resulting from an affine invariant theory must satisfy the differential equation 
\begin{equation} \label{eq22}
x^{\mu}\partial_{\nu}G(x)=i[Q^{\mu}_{\;\;\nu},G(x)],
\end{equation}
where $Q^{\mu}_{\;\;\nu}$ corresponds to whatever representation $G(x)$ is. For a scalar propagator, the only solution is $G(x)=\delta^{d}(x)$.\footnote{Really this is for a scalar \textit{density} field $\Phi'(x')=\sqrt{|\det{M^{-1}}|}\Phi(M^{-1}x')$.} For other finite dimensional representations, the solution will always have at most point support. 

For infinite dimensional representations, the solution to equation (\ref{eq22}) is no longer obvious. It is likely though that for the representation used in section \ref{sec4}, such a propagator will have finite dimensional Lorentz components which are covariant only under the Lorentz subgroup, and mix nontrivially among themselves under the other group transformations. This is in fact what typically happens in a conformal field theory, whose field representations are generically infinite. This affords the possibility of having a nontrivial propagator, and correspondingly nontrivial dynamics which could exhibit spontaneous symmetry breaking.

To actually exhibit spontaneous symmetry breaking, we need to incorporate interactions somehow. This can be done in a number of ways. A first instinct one might have is to couple this affine Dirac field $\Psi$ to some gauge field $A_{\mu}$ with some gauge group $G$ (e.g. $SU(N)$ or $U(N)$) by replacing $\partial_{\mu}$ with $D_{\mu}=\partial_{\mu}-iA_{\mu}$. Here however the standard Yang-Mills kinetic term $-\frac{1}{4g^2}\text{tr}(F^{\mu\nu}F_{\mu\nu})$ cannot be written down, since this requires a metric. Coupling $A_{\mu}$ to $\Psi$ in this way can at most be interpreted as the strong coupling limit $g\to\infty$ of the Yang-Mills theory. In this limit, $A_{\mu}$ acts as an auxiliary field which constrains the current $J_{\alpha}^{\mu}=(\Psi,\Gamma^{\mu}t_{\alpha}\Psi)$ to vanish.

Another possibility is to model interactions with a simple four fermion term, in analogy with the Nambu-Jona-Lasinio (NJL) Model \cite{PhysRev.122.345}. We will write this action in the form
\begin{align} \label{eq23}
S=-\int d^dx (\Psi,\Gamma^{\mu}\partial_{\mu}\Psi)-\int d^dx\sum_{nm}\Big(&g_n(\Psi(x),A^{\;\;m}_{n}\Psi(x))(\Psi(x),B_{mn}\Psi(x))\Big),
\end{align} 
where $A^{\;\;m}_n$ and $B_{mn}$ are at this stage generic constant operators. The current representation we are using allows for an infinite number of possible $A$'s and $B$'s, e.g. $A=\Gamma^{\mu_1}\cdots\Gamma^{\mu_n}$ and $B=\Omega_{\mu_1}\cdots\Omega_{\mu_n}$. There is no good reason to choose a particular set of $A$'s and $B$'s, so we leave it arbitrary. This representation does not in general offer operators $A^{\;\;m}_n$ and $B_{mn}$ which leave the action invariant under affine transformations, but only special affine transformations, so this model in general will only allow us to study the symmetry breaking pattern $SL(d,\mathbb{R})\to H$. This is not much of a restriction, since the result in section \ref{sec2} is not changed dramatically if $GL(d,\mathbb{R})$ is replaced with $SL(d,\mathbb{R})$.

This theory is analogous to the NJL model, which was originally used to model chiral symmetry breaking via dynamical mass generation. Because there is no chiral invariant notion of mass, a dynamical generation of mass will spontaneously break the symmetry. Similarly, there is no affine invariant notion of mass, and so any generation of some well-defined notion of mass will break the symmetry. Peculiarly, while there is no affine invariant notion of mass, the representation we use does allow for an affine invariant ``mass term" $m(\Psi,\Psi)$ in the action. The effect of this term cannot be interpreted as adding a mass to the field $\Psi$.\footnote{This term avoids breaking affine invariance by not adding a single mass, but a continuous spectrum of masses of all values. This is indeed what you would expect to happen, since a continuous affine transformation does not leave the equation $p^2=-m^2$ invariant.} Instead, if a dynamical generation of a term like $m\eta_{\mu\nu}(\Psi,\Gamma^{\mu}\Gamma^{\nu}\Psi)$ or $m\eta^{\mu\nu}(\Psi,\Omega_{\mu}\Omega_{\nu}\Psi)$ occurs, this would spell the symmetry breakdown we are looking for. 

In order to study the symmetry breaking aspects of these interacting theories, one should first develop a formalism for perturbative calculations which respects the affine symmetry. Feynman diagrams are optimally designed to calculate Lorentz invariant perturbations from free field theory. When a theory has an extended spacetime symmetry, a Feynman diagram perturbation expansion is no longer optimal, and may in fact obscure deep results that a formalism which respects the full symmetry would make obvious. This is true, for example, in the case of supersymmetry, where Feynman diagram calculations yielded surprising cancellations, which only after a supergraph formalism was developed became less surprising \cite{GRISARU1979429}. We do not attempt to develop an affine invariant formalism for perturbative calculations in this paper, and leave it for future work. A first step would be to determine the exact free propagator of the Dirac action (\ref{eq21}) (see appendix \ref{appC} for progress in this direction).

\section{Conclusion\label{sec6}}
\setcounter{equation}{0}
\setcounter{Roman}{0}

The possibility of a dynamically generated metric has motivated this search for affine invariant theories which spontaneously break down to the Lorentz subgroup. This has led us to use an infinite dimensional spinor representation $\Psi$ 
of $GL(d,\mathbb{R})$ to construct an affine generalization of the Dirac action.  We next showed how a large class of affine invariant interactions can be constructed to obtain interacting theories. Explicit constructions were presented in $d=3$. 
The constructions being non-trivial, they provide an essential toolbox for obtaining such actions to study their potential
spontaneous symmetry breaking. 
These theories, which a priori have no notion of distance, have the ability to produce a metric via the expectation value $\langle (\Psi,\Gamma^{\mu}\Gamma^{\nu}\Psi)\rangle\sim \bar{g}^{\mu\nu}$. Should such a non-vanishing expectation value form, the model's (special) affine symmetry would spontaneously break down to one of the subgroups $SO(p,q)$, $p+q=d$. Which of the subgroups $SO(p,q)$, which in the unbroken phase are all on equal footing, is actually picked out is determined by the signature of the  matrix $\bar{g}^{\mu\nu}$, which in turn is determined by the form of the interactions. If this signature is Lorentzian, this would amount to a dynamical explanation of the observed signature of spacetime. 
GR then emerges as the long distance effective theory in the manner detailed in the cited literature, and reviewed in section \ref{sec2} above, with the 
curved metric $g_{\mu\nu}(x)$ being an appropriate nonlinear combination of the resulting Goldstone fields. 
As already pointed out above, a good deal of work is needed for the further development of this program to which we hope to return in the future.  

\setcounter{equation}{0}
\appendix
\renewcommand{\theequation}{\mbox{\Alph{section}.\arabic{equation}}}

\section*{Appendices} 

\section{Unitary Representations of $\overline{SL(4,\mathbb{R})}$ \label{appA}}
 For noncompact Lie groups, there are no nontrivial finite dimensional faithful unitary representations. The infinite dimensional unitary representations of $\overline{SL(4,\mathbb{R})}$ can be defined in the basis of its maximally compact subgroup $\overline{SO(4)}=SU(2)\otimes SU(2)$. Restricting ourselves to multiplicity-free representations, the homogeneous vector space can be taken to be the set of all $|j_1 m_1;j_2 m_2\rangle $ with $j_1,j_2=0,\frac{1}{2},1,\frac{3}{2},\dots $ and $|m_i|\leq j_i$.

The $SU(2)\otimes SU(2)$ subgroup is generated by 
\begin{equation} \label{eqA1}
J_i^{(1)}=\frac{1}{4}\epsilon_{ijk}J_{jk}+\frac{1}{2}T_{0i},\quad\quad J_i^{(2)}=\frac{1}{4}\epsilon_{ijk}J_{jk}-\frac{1}{2}T_{0i},
\end{equation}
with $i,j,k$ denoting spatial indices. The rest of the nine generators form an $SU(2)\otimes SU(2)$ $(1,1)$ tensor operator $Z_{\alpha\beta}$, with $\alpha,\beta=0,\pm 1$. In terms of these generators, the $\overline{SL(4,\mathbb{R})}$ algebra reads:
\begin{align} \label{eqA2:1}
&[J_0^{(p)},J_{\pm}^{(q)}]=\pm\delta_{pq}J_{\pm}^{(p)}, \quad p,q=1,2,\\ \label{eqA2:2}
&[J_{+}^{(p)},J_{-}^{(q)}]=2\delta_{pq}J_0^{(p)},\\ \label{eqA2:3}
&[J_{0}^{(1)},Z_{\alpha\beta}]=\alpha Z_{\alpha\beta},\\ \label{eqA2:4}
&[J_{0}^{(2)},Z_{\alpha\beta}]=\beta Z_{\alpha\beta},\\ \label{eqA2:5}
&[J_{\pm}^{(1)},Z_{\alpha\beta}]=\sqrt{2-\alpha(\alpha\pm 1)} Z_{\alpha\pm 1\beta},\\ \label{eqA2:6}
&[J_{\pm}^{(2)},Z_{\alpha\beta}]=\sqrt{2-\beta(\beta\pm 1)} Z_{\alpha\beta\pm 1},\\ \label{eqA2:7}
&[Z_{11},Z_{-1-1}]=-(J_0^{(1)}+J_0^{(2)}).
\end{align}
The remaining commutation relations can be determined from these. The $J_{0}^{(p)}$, $J_{\pm}^{(p)}$ have the well-known action on the $|j_1 m_1;j_2 m_2\rangle $ vectors, and the general solution to the matrix elements of $Z_{\alpha\beta}$ are \cite{doi:10.1063/1.526758}
\begin{align}\langle j_{1}'m_1';j_2' m_2' |Z_{\alpha\beta}&|j_1m_1;j_2m_2\rangle=\nonumber\\&-i(-1)^{2(j_1'+j_2')-m_1'-m_2'}\resizebox{0.4\hsize}{!}{$\sqrt{(2j_1'+1)(2j_2'+1)(2j_1+1)(2j_2+1)}$}\nonumber\\&\times(e-\frac{1}{2}[j_1'(j_1'+1)+j_2'(j_2'+1)-j_1(j_1+1)-j_2(j_2+1)])\nonumber\\ \label{eqA3} &\times{\textstyle\bigl(\begin{smallmatrix}j_1' & 1 & j_1 \\ -m_1' & \alpha & m_1\end{smallmatrix}\bigr)\bigl(\begin{smallmatrix}j_2' & 1 & j_2 \\ -m_2' & \beta & m_2\end{smallmatrix}\bigr)\bigl( \begin{smallmatrix}j_1' & 1 & j_1 \\ 0 & 0 & 0\end{smallmatrix}\bigr)\bigl(\begin{smallmatrix}j_2' & 1 & j_2 \\ 0 & 0 & 0\end{smallmatrix}\bigr)},\end{align}
where $e$ is the Casimir label associated with $Q^{\alpha}_{\,\,\beta}Q^{\beta}_{\,\,\alpha}=\frac{1}{4}e^2-4$, the $\bigl(\begin{smallmatrix}j_i' & 1 & j_i \\ -m_i' & x & m_i\end{smallmatrix}\bigr)$ are $3$-$j$ symbols, and when the $j_i$'s are half-integers, $ \bigl(\begin{smallmatrix}j_i' & 1 & j_i \\ 0 & 0 & 0\end{smallmatrix}\bigr)$ is understood by taking the corresponding expression for integer values and continuing it to half-integer ones. Studying the $3$-$j$ symbols, we see that $Z$ can change the $SU(2)$ Casimir labels with transitions $(j_1',j_2')=(j_1\pm 1,j_2\pm 1)$ and $(j_1',j_2')=(j_1\pm 1,j_2\mp 1)$. In general then, there are eight invariant $(j_1,j_2)$ lattices, generated from the points $(0,0)$, $(\frac{1}{2},\frac{1}{2})$, $(0,1)$, $(\frac{1}{2},\frac{3}{2})$, $(\frac{1}{2},0)$, $(0,\frac{1}{2})$, $(\frac{3}{2},0)$, and $(0,\frac{3}{2})$. To get the spinorial representations of $\overline{SL(4,\mathbb{R})}$, we must choose a representation which acts on one of the latter four lattices. We choose the $(0,\frac{1}{2})$ lattice since this includes the lowest weight spinor representations (the $(\frac{1}{2},0)$ lattice is related by a 4-dimensional parity transformation).

Unitarity of a representation is a statement about the Hermiticity of the generators with respect to its Hilbert space. Thus we must find an inner product $(f,g)$ between vectors on the $(0,\frac{1}{2})$ lattice that render the generators Hermitian with respect to it. The most general inner product on this vector space has some kernel $\kappa (j_1,j_2)$ to be determined:
\begin{equation} \label{eqA4}
(f,g)=\sum_{j_1m_1j_2m_2}f^{*j_1j_2}_{\,m_1m_2}\kappa(j_1,j_2)g^{j_1j_2}_{m_1m_2},
\end{equation}
the $J_i^{(p)}$ generators are already Hermitian with respect to this inner product. The condition of Hermiticity for the shear generators in terms of $Z$ is $Z^{\dagger}_{\alpha\beta}=(-1)^{\alpha-\beta}Z_{-\alpha-\beta}$. In order for this to hold, the kernel must satisfy
\begin{align} \label{eqA5} (e-\frac{1}{2}&[j_1'(j_1'+1)+j_2'(j_2'+1)-j_1(j_1+1)-j_2(j_2+1)])\kappa(j_1',j_2')\\&=(-e^*-\frac{1}{2}[j_1'(j_1'+1)+j_2'(j_2'+1)-j_1(j_1+1)-j_2(j_2+1)])\kappa(j_1,j_2).\nonumber\end{align}

We now state the value of $e$ and $\kappa(j_1,j_2)$ for the representation we use in our work.
\begin{equation} \label{eqA6}
e=\frac{1}{2},\quad\quad\kappa(j_1,j_2)=\frac{(j_1+j_2+\frac{1}{2})!!(j_2-j_1-\frac{1}{2})!!}{(j_1+j_2-\frac{1}{2})!!(j_2-j_1-\frac{3}{2})!!}.
\end{equation}

With this value of $e$, we have $\langle j+1,j-\frac{1}{2}|Z|j,j+\frac{1}{2}\rangle=0$. This implies the invariance of a sublattice consisting of points $(j_1,j_2)=(2n,2m+\frac{1}{2}),(2n+1,2m+1+\frac{1}{2})$ with $m,n$ non-negative integers and $j_2\geq j_1+\frac{1}{2}$. This is the only multiplicity free unitary representation acting on the $(0,\frac{1}{2})$ lattice. 

Other combinations of $e$ and choice of lattice will result in a different kernel $\kappa(j_1,j_2)$, provided of course one can satisfy equation (\ref{eqA5}). We will not outline the rest of the unitary representations.

\section{Defining Objects Used in the Construction of the 3-dimensional Action \label{appB}}
\setcounter{equation}{0}
The objects used to construct the 3-dimensional affine Dirac theory are quite complicated, so we will spend some time here to describe them. 

The spinor matrix tensor structures $\Pi^{\mu_1\cdots\mu_s\nu_1\cdots\nu_s}$, $\Pi^{\mu_1\cdots\mu_s\lambda\nu_1\cdots\nu_s}$ are all totally symmetric in their $\mu$ and $\nu$ indices separately, as well as gamma traceless $\gamma_{\mu_1}\Pi^{\mu_1\cdots\mu_s\nu_1\cdots\nu_s}=0$, $\Pi^{\mu_1\cdots\mu_s\nu_1\cdots\nu_s}\gamma_{\nu_1}=0$ etc. Their adjoints are $\big(\Pi^{\lambda}_s\big)^{\dagger}=\gamma_0\Pi^{\lambda}_s\gamma_0$, $\big(\Pi_s\big)^{\dagger}=-\gamma_0\Pi_s\gamma_0$, and they have the normalizations $\Pi_s\Pi_s=\Pi_s$, $\Pi_{s\lambda}\Pi^{\lambda}_s=\frac{2s+3}{2s+1}\Pi_s$. Here are the first few spinor matrices:
\pagebreak
\begin{align} \label{eqB1:1} &\Pi=1,\\ \label{eqB1:2} &\Pi^{\lambda}=\gamma^{\lambda},\\ \label{eqB1:3} & \Pi^{\mu\nu}=\frac{1}{3}\big(2\eta^{\mu\nu}-\gamma^{\mu\nu}\big),\\ \label{eqB1:4} & \Pi^{\mu\lambda\nu}=\frac{1}{9}\big(4\eta^{\mu\nu}\gamma^{\lambda}-\eta^{\lambda\mu}\gamma^{\nu}-\eta^{\lambda\nu}\gamma^{\mu}+5\gamma^{\mu\lambda\nu}\big),\\
 \label{eqB1:5}
&\Pi^{\mu_1\mu_2\nu_1\nu_2}=\frac{1}{10}\big(-2\eta^{\mu_1\mu_2}\eta^{\nu_1\nu_2}+3\eta^{\mu_1\nu_1}\eta^{\mu_2\nu_2}+3\eta^{\mu_1\nu_2}\eta^{\mu_2\nu_1}\nonumber\\&-\eta^{\mu_1\nu_1}\gamma^{\mu_2\nu_2}-\eta^{\mu_2\nu_1}\gamma^{\mu_1\nu_2}-\eta^{\mu_1\nu_2}\gamma^{\mu_2\nu_1}-\eta^{\mu_2\nu_2}\gamma^{\mu_1\nu_1}\big),\\ \label{eqB1:6}
&\Pi^{\mu_1\mu_2\lambda\nu_1\nu_2}=\frac{1}{100}\big((-20\eta^{\mu_1\mu_2}\eta^{\nu_1\nu_2}+22\eta^{\mu_1\nu_1}\eta^{\mu_2\nu_2}+22\eta^{\mu_2\nu_1}\eta^{\mu_1\nu_2})\gamma^{\lambda}\nonumber\\&+(8\eta^{\lambda\mu_2}\eta^{\nu_1\nu_2}-6\eta^{\lambda\nu_2}\eta^{\nu_1\mu_2}-6\eta^{\lambda\nu_1}\eta^{\nu_2\mu_2})\gamma^{\mu_1}+(8\eta^{\lambda\mu_1}\eta^{\nu_1\nu_2}-6\eta^{\lambda\nu_2}\eta^{\nu_1\mu_1}-6\eta^{\lambda\nu_1}\eta^{\nu_2\mu_1})\gamma^{\mu_2}\nonumber\\&+(8\eta^{\mu_1\mu_2}\eta^{\lambda\nu_2}-6\eta^{\mu_1\nu_2}\eta^{\lambda\mu_2}-6\eta^{\mu_2\nu_2}\eta^{\lambda\mu_1})\gamma^{\nu_1}+(8\eta^{\mu_1\mu_2}\eta^{\lambda\nu_1}-6\eta^{\mu_1\nu_1}\eta^{\lambda\mu_2}-6\eta^{\mu_2\nu_1}\eta^{\lambda\mu_1})\gamma^{\nu_2}\nonumber\\&-14(\gamma^{\mu_1\nu_1\lambda}\eta^{\mu_2\nu_2}+\gamma^{\mu_2\nu_1\lambda}\eta^{\mu_1\nu_2}+\gamma^{\mu_1\nu_2\lambda}\eta^{\mu_2\nu_1}+\gamma^{\mu_2\nu_2\lambda}\eta^{\mu_1\nu_1})\big).
\end{align}

The even rank matrices $\Pi^{\mu_1\cdots\mu_s\nu_1\cdots\nu_s}$ are projection matrices ($\Pi^2=\Pi$) which leave the subspace of totally symmetric and gamma traceless rank $s$ spinor fields invariant, while the odd rank matrices $\Pi^{\mu_1\cdots\mu_s\lambda\nu_1\cdots\nu_s}$ behave like vector operators on this subspace (one can think of them as a spin $s+\frac{1}{2}$ generalization of $\gamma^{\lambda}$ in three dimensions). In this tensor basis these objects are very complicated, but in the $SO(2,1)$ spin basis $|s+\frac{1}{2},m\rangle$ they're defined in, they have a very simple form
\begin{equation} \label{eqB2}
\Pi_s|s+\frac{1}{2},m\rangle=|s+\frac{1}{2},m\rangle,\quad \Pi^0_s|s+\frac{1}{2},m\rangle=i\frac{m}{s+\frac{1}{2}}|s+\frac{1}{2},m\rangle,
\end{equation}
where $\Pi_s$ are the even rank spinor matrices and $\Pi^0_s$ is the $0$th component of the vector operator $\Pi^{\lambda}_s$ which defines the odd rank spinor matrices action in this basis. If one wants to find these matrices in the tensor basis for some $s$, they can use equations (\ref{eqB2}) and simply change the basis.

The matrices $\Gamma^{s,s}$, $\Gamma^{s,s-1}$, and $\Gamma^{s,s+1}$ are infinite dimensional matrices acting on the $n,h$ labels. $\Gamma^{s,s}$ is Hermitian with respect to the affine invariant inner product, and $(\Gamma^{s,s+1})^{\dagger}=\Gamma^{s+1,s}$. Explicitly,
\pagebreak
\begin{align}&(\Gamma^{s,s})_{nh,n'h'}=\textstyle{\frac{1}{2(2s+3)}}\Big(\resizebox{.15\hsize}{!}{$(4n+3)(4h+1)$}\;\delta_{nn'}\delta_{hh'}\nonumber\\&+\resizebox{.1\hsize}{!}{$2n(4h+1)$}\sqrt{\textstyle{\frac{(2n-1-s)(2n-s)(2n+1+s)(2n+2+s)}{(2n+2h)(2n+2h+2)(2n-2h-1)(2n-2h+1)}}}\;\delta_{n-1,n'}\delta_{hh'}\nonumber\\&+\resizebox{.15\hsize}{!}{$(2n+3)(4h+1)$}\sqrt{\textstyle{\frac{(2n+1-s)(2n+2-s)(2n+3+s)(2n+4+s)}{(2n+2h+2)(2n+2h+4)(2n-2h+1)(2n-2h+3)}}}\;\delta_{n+1,n'}\delta_{hh'}\nonumber\\&-\resizebox{.15\hsize}{!}{$(4n+3)(2h-1)$}\sqrt{\textstyle{\frac{(s+2-2h)(s+1-2h)(s+2h)(s+1+2h)}{(2n+2h)(2n+2h+2)(2n-2h+1)(2n-2h+3)}}}\;\delta_{nn'}\delta_{h-1,h'}\nonumber\\&-\resizebox{.15\hsize}{!}{$(4n+3)(2h+2)$}\sqrt{\textstyle{\frac{(s-2h)(s-1-2h)(s+2+2h)(s+3+2h)}{(2n+2h+2)(2n+2h+4)(2n-2h-1)(2n-2h+1)}}}\;\delta_{nn'}\delta_{h+1,h'}\Big),
\nonumber\\&(\Gamma^{s,s+1})_{nh,n'h'}=\textstyle{\frac{1}{\sqrt{(s+2)(2s+3)}}}\Big(\resizebox{.42\hsize}{!}{$\sqrt{(s+1-2h)(s+2+2h)(2n+3+s)(2n-s)}$}\;\delta_{nn'}\delta_{hh'}\nonumber\\&+\textstyle{\frac{1}{2}}\resizebox{.25\hsize}{!}{$2n\sqrt{(s+1-2h)(s+2+2h)}$}\sqrt{\textstyle{\frac{(2n-2-s)(2n-1-s)(2n-s)(2n+2+s)}{(2n+2h)(2n+2h+2)(2n-2h-1)(2n-2h+1)}}}\;\delta_{n-1,n'}\delta_{hh'}\nonumber\\&+\textstyle{\frac{1}{2}}\resizebox{.32\hsize}{!}{$(2n+3)\sqrt{(s+1-2h)(s+2+2h)}$}\sqrt{\textstyle{\frac{(2n+3+s)(2n+4+s)(2n+5+s)(2n+1-s)}{(2n+2h+2)(2n+2h+4)(2n-2h+1)(2n-2h+3)}}}\;\delta_{n+1,n'}\delta_{hh'}\nonumber\\&-\textstyle{\frac{1}{2}}\resizebox{.3\hsize}{!}{$(2h-1)\sqrt{(2n+3+s)(2n-s)}$}\sqrt{\textstyle{\frac{(s+1-2h)(s+2-2h)(s+3-2h)(s+1+2h)}{(2n+2h)(2n+2h+2)(2n-2h+1)(2n-2h+3)}}}\;\delta_{nn'}\delta_{h-1,h'}\nonumber\\&+\textstyle{\frac{1}{2}}\resizebox{.3\hsize}{!}{$(2h+2)\sqrt{(2n+3+s)(2n-s)}$}\sqrt{\textstyle{\frac{(s-2h)(s+2+2h)(s+3+2h)(s+4+2h)}{(2n+2h+2)(2n+2h+4)(2n-2h-1)(2n-2h+1)}}}\;\delta_{nn'}\delta_{h+1,h'}\Big),\nonumber\\ \label{eqB3} &(\Gamma^{s+1,s})=(\Gamma^{s,s+1})^{\dagger}.
\end{align}

Throughout this paper there are also several mentions of the covariant vector operator $\Omega_{\mu}\equiv Q^{d+1}_{\;\;\;\;\;\;\mu}$. In this 3-dimensional representation, $\Omega_{\mu}$ is related to the contravariant vector $\Gamma^{\mu}$ by
\begin{equation} \label{eqB4}
\Omega_{\mu}=\eta_{\mu\nu}\tilde{I}\Gamma^{\nu}\tilde{I}
\end{equation}

Where $\tilde{I}$ is a simple matrix in the $n,h$ labels $(\tilde{I})^{s,s'}_{nh,n'h'}=(-1)^{n+h}\delta_{nn'}\delta_{hh'}\delta_{s,s'}$, where the superscript $s,s'$ is used to denote the component of  the matrix which sends a tensor rank $s'$ Lorentz field to a tensor rank $s$ Lorentz field.

\section{Finding the Propagator \label{appC}}
\setcounter{Roman}{0}
\setcounter{equation}{0}
In order to perform any calculation in the theory described in section \ref{sec5} in a way which preserves $SL(d,\mathbb{R})$, one needs at least the free propagator. The infinite dimensional nature of the representation makes finding the exact propagator a technical challenge. In the 3-dimensional example, a calculation of the propagator $(\Gamma^{\mu}p_{\mu})^{-1}$ requires finding the inverses of all $\Pi^{\mu_1\cdots\mu_s\lambda\nu_1\cdots\nu_s}p_{\lambda}$'s and $\Gamma^{s,s}$'s. Whereas finding the inverses of $\Pi^{\mu_1\cdots\mu_s\lambda\nu_1\cdots\nu_s}p_{\lambda}$ are pretty straightforward, the inverses of the infinite dimensional matrices $\Gamma^{s,s}$ are nontrivial. $\Gamma^{0,0}$ and $\Gamma^{1,1}$ have the following inverses
\begin{equation}
(\Gamma^{0,0})^{-1}_{n,n'}=6(-1)^{n+n'}{\textstyle\frac{(2\text{max}(n,n'))!!}{(2\text{max}(n,n')+1)!!}\frac{(2n'+1)!!}{(2n')!!}},
\end{equation}
\begin{equation}
(\Gamma^{1,1})^{-1}_{n,n'}=2(-1)^{n+n'}{\textstyle\frac{(2\text{max}(n,n'))!!}{(2\text{max}(n,n')+1)!!}\frac{(2n'+1)!!}{(2n')!!}\sqrt{\frac{2\text{min}(n,n')(2\text{min}(n,n')+3)}{2\text{max}(n,n')(2\text{max}(n,n')+3)}}}.
\end{equation}

These inverses were obtained because $\Gamma^{0,0}$ and $\Gamma^{1,1}$ have a relatively simple form, i.e. they are both tridiagonal. There is a beautiful formula for the inverse of a nonsingular tridiagonal matrix $T$
$$T=\begin{pmatrix}
a_1 & b_1 & 0 & \cdots & 0\\
c_1 & a_2 & b_2  & \ddots&\vdots \\
0 & c_2 & \ddots & \ddots & 0 \\
\vdots & \ddots & \ddots & \ddots & b_{n-1}\\
0& \cdots&0 & c_{n-1} & a_n
\end{pmatrix}
$$
given by
$$(T^{-1})_{ij}=
\begin{cases}
(-1)^{i+j}b_i\cdots b_{j-1}\theta_{i-1}\phi_{j+1}/\theta_n & i<j\\
\theta_{i-1}\phi_{i+1}/\theta_n & i=j\\
(-1)^{i+j}c_j\cdots c_{i-1}\theta_{j-1}\phi_{i+1}/\theta_n & i>j
\end{cases}$$
where the $\theta_i$ satisfy the recursive relation 
$$\theta_i=a_i\theta_{i-1}-b_{i-1}c_{i-1}\theta_{i-2}\quad i=2,3,\dots, n$$
with initial conditions $\theta_0=1$, $\theta_1=a_1$ and the $\phi_i$ satisfy 
$$\phi_i=a_i\phi_{i+1}-b_ic_i\phi_{i+2}\quad i=n-1,\dots,1$$
with initial conditions $\phi_{n+1}=1$, $\phi_n=a_n$. Note that the sequences $\theta_i$ and $\phi_i$ are determinants of the submatrices
$$\theta_i=\begin{vmatrix}
a_1 & b_1 & 0 & \cdots & 0\\
c_1 & a_2 & b_2  & \ddots&\vdots \\
0 & c_2 & \ddots & \ddots & 0 \\
\vdots & \ddots & \ddots & \ddots & b_{i-1}\\
0& \cdots&0 & c_{i-1} & a_i
\end{vmatrix},\quad \phi_i=\begin{vmatrix}
a_i & b_i & 0 & \cdots & 0\\
c_i & a_{i+1} & b_{i+1}  & \ddots&\vdots \\
0 & c_{i+1} & \ddots & \ddots & 0 \\
\vdots & \ddots & \ddots & \ddots & b_{n-1}\\
0& \cdots&0 & c_{n-1} & a_n
\end{vmatrix}.$$

For the cases of $\Gamma^{0,0}$, $\Gamma^{1,1}$, $\theta_i$ and $\phi_i$ can be explicitly solved for. This formula stops being useful for $\Gamma^{s,s}$ with $s>1$, as the rest of these are no longer tridiagonal, but \textit{block} tridiagonal. Such an $mn\times mn$ matrix can be written in the form 
$$T=\begin{pmatrix}
A_1 & B_1 & 0 & \cdots & 0\\
C_1 & A_2 & B_2  & \ddots&\vdots \\
0 & C_2 & \ddots & \ddots & 0 \\
\vdots & \ddots & \ddots & \ddots & B_{n-1}\\
0& \cdots&0 & C_{n-1} & A_n
\end{pmatrix}
$$
where now $A_i$, $B_i$ and $C_i$ are $m\times m$ square matrices ($(\lfloor\frac{s}{2}\rfloor+1)\times(\lfloor\frac{s}{2}\rfloor+1)$ in the case of $\Gamma^{s,s}$). We would like to develop a generalization of the formula for the inverse $m\times m$ block elements $(T^{-1})_{ij}$ such that
$$T^{-1}=\begin{pmatrix}
(T^{-1})_{11} &(T^{-1})_{12} & (T^{-1})_{13} & \cdots & (T^{-1})_{1n}\\
(T^{-1})_{21} & (T^{-1})_{22} & (T^{-1})_{23}  & \ddots&\vdots \\
(T^{-1})_{31} &(T^{-1})_{32} & \ddots & \ddots & (T^{-1})_{n-2n} \\
\vdots & \ddots & \ddots & \ddots &(T^{-1})_{n-1n}\\
(T^{-1})_{n1}& \cdots&(T^{-1})_{nn-2} & (T^{-1})_{nn-1} &(T^{-1})_{nn}
\end{pmatrix}.$$

To do this we first note the formula for the inverse of a block matrix partitioned into four parts
$$\begin{pmatrix}
A & B\\
C & D
\end{pmatrix}.$$
If we assume that both $A$ and $D$ are nonsingular, the inverse may be written 
$$\begin{pmatrix}
A & B\\
C & D
\end{pmatrix}^{-1}=\begin{pmatrix}
(A-BD^{-1}C)^{-1} & -A^{-1}B(D-CA^{-1}B)^{-1}\\
-(D-CA^{-1}B)^{-1}CA^{-1} & (D-CA^{-1}B)^{-1}
\end{pmatrix}.$$
Using this, a formula for $(T^{-1})_{ij}$ can be obtained:
\begin{equation} \label{eqC:1} (T^{-1})_{ij}=
\begin{cases}
(-1)^{i+j}\Theta_i^{-1}B_i\cdots\Theta_{j-1}^{-1}B_{j-1}(T^{-1})_{jj}& i<j\\
(A_i-C_{i-1}\Theta_{i-1}^{-1}B_{i-1}-B_i\Phi_{i+1}^{-1}C_i)^{-1} & i=j\\
(-1)^{i+j}(T^{-1})_{ii}C_{i-1}\Theta^{-1}_{i-1}\cdots C_j\Theta_{j}^{-1} & i>j
\end{cases}\end{equation}
where the $\Theta_i$ satisfy the recursive relation
$$\Theta_i=A_i-C_{i-1}\Theta_{i-1}^{-1}B_{i-1}\quad i=2,3,\dots,n$$
with initial condition $\Theta_1=A_1$ and the $\Phi_i$ satisfy
$$\Phi_i=A_i-B_i\Phi_{i+1}^{-1}C_i\quad i=n-1,\dots,1$$
with initial condition $\Phi_n=A_n$. This formula of course presumes the existence of all the inverses indicated above. The usefulness of this formula has yet to be seen, but it seems to be as close as one can get to an analogous formula to the tridiagonal case. It may be useful in writing down the inverse of the full affine propagator $\Gamma^{\mu}p_{\mu}$, as this itself is block tridiagonal, with $A_s=\Gamma^{s,s}\Pi^{\mu_1\cdots\mu_s\lambda\nu_1\cdots\nu_s}p_{\lambda}$, $B_s=-\Gamma^{s,s-1}\Pi^{\mu_1\cdots\mu_s\lambda\nu_1\cdots\nu_{s-1}}p_{\lambda}$ and ${C_s=-\Gamma^{s,s+1}\Pi^{\mu_1\cdots\mu_s\lambda\nu_1\cdots\nu_{s+1}}p_{\lambda}}$.

For $s>1$, the inverses of $\Gamma^{s,s}$ have yet to be determined. At this stage, we are therefore able to explicitly write down the propagator if one only includes the fields $\psi^n$ and $\psi^n_{\mu}$:
\begin{align}
&\langle \psi^n(x)\overline{\psi^{n'}}(y)\rangle=-\frac{6}{9}(-1)^{n+n'}{\textstyle\frac{(2\text{max}(n,n'))!!}{(2\text{max}(n,n')+1)!!}\frac{(2n'+1)!!}{(2n')!!}}\int\frac{d^3p}{(2\pi)^3}\frac{-\fsl{p}}{p^2-i\varepsilon}e^{ip\cdot(x-y)},\\
&\langle \psi^n(x)\overline{\psi^{n'}}_{\mu}(y)\rangle=-\frac{5}{\sqrt{3}}(-1)^{n+n'}{\textstyle\frac{2\text{max}(n'-n,0)}{\sqrt{2n'(2n'+3)}}} \int\frac{d^3p}{(2\pi)^3}\frac{p^{\lambda}\Pi_{\lambda\mu}}{p^2-i\varepsilon}e^{ip\cdot(x-y)},
\end{align}

\pagebreak
\begin{align}
\langle \psi^n_{\mu}(x)\overline{\psi^{n'}}_{\nu}(y)\rangle=&18(-1)^{n+n'}\textstyle{\frac{(2\text{max}(n,n'))!!}{(2\text{max}(n,n')+1)!!}\frac{(2n'+1)!!}{(2n')!!}\sqrt{\frac{2\text{min}(n,n')(2\text{min}(n,n')+3)}{2\text{max}(n,n')(2\text{max}(n,n')+3)}}}\nonumber\\&\times\int\frac{d^3p}{(2\pi)^3}\frac{1}{p^2-i\varepsilon}\Big(p^{\lambda}\Pi_{\mu\lambda\nu}-(\eta_{\mu\nu}-\frac{p_{\mu}p_{\nu}}{p^2})\fsl{p}-p^{\lambda}\gamma_{\mu\lambda\nu}\Big)e^{ip\cdot(x-y)}.
\end{align}

Of course, any calculation done with a propagator which includes only spin $1/2$ and $3/2$ fields explicitly breaks affine invariance, and so nothing regarding the theory's spontaneous breaking can be determined. The above propagator does however have properties that one might expect from an affine propagator. Namely, that if we add a ``mass'' term $m(\Psi,\Psi)$ to the action, the above propagator changes such that it has a continuum of masses. This is an expected feature of any affine invariant theory, since a continuous affine transformation does not leave the equation $p^2=-m^2$ invariant. This continuum of masses follows from the fact that $\Gamma^{0,0}$ and $\Gamma^{1,1}$ have continuous spectra. Indeed, $\Gamma^{0,0}$ has eigenvectors
\begin{equation}\Gamma^{0,0}v(\lambda)=\lambda v(\lambda),\end{equation}
\begin{equation}v^n(\lambda)=\Big(\frac{3}{4\pi\lambda}\Big)^{1/4}e^{-3\lambda/2}\frac{1}{2^n(2n+1)!!}H_{2n+1}(\sqrt{3\lambda}),\end{equation}
where $H_{2n+1}(x)$ are the standard odd degree Hermite polynomials. This expression works formally for all complex $\lambda$, but it is only for $\lambda>0$ that we get the following orthonormality conditions
$$(v(x),v(y))=\sum_{n=0}^{\infty}\frac{(2n+1)!!}{(2n)!!}v^{n*}(x)v^n(y)=\delta(x-y),$$
$$\int_{0}^{\infty}dx v^n(x)v^{n'*}(x)\frac{(2n'+1)!!}{(2n')!!}=\delta_{nn'},$$
and $\frac{1}{\lambda\sqrt{5}}\Gamma^{1,0}v(5\lambda/3)$ is an eigenvector of $\Gamma^{1,1}$ with eigenvalue $\lambda$. For simplicity, we write the propagator with the above mass term just including the $\psi^n$ fields
\begin{align}&\langle\psi^n(x)\overline{\psi^{n'}}(y)\rangle=\int\frac{d^3p}{(2\pi)^3}\int_0^{\infty}d\mu\rho^{nn'}(\mu)\frac{-\fsl{p}+i\mu}{p^2+\mu^2-i\varepsilon}e^{ip\cdot(x-y)},\end{align}
\begin{equation}\rho^{nn'}(\mu)=\sqrt{\frac{3}{4\pi m\mu}}e^{-3m/\mu}\frac{H_{2n+1}(\sqrt{3m/\mu})H_{2n'+1}(\sqrt{3m/\mu})}{2^{n+n'}(2n+1)!!(2n')!!}.\end{equation}

Notice that if the inverse of $\Gamma^{\mu}p_{\mu}$ exists, it is difficult to see how the recursive formula (\ref{eqC:1}) produces a continuum of masses, and will only have poles at $p^2=0$.

\end{document}